# Simulating Space and Time[1]

*Brian Whitworth*
Massey University, Albany, Auckland, New Zealand

*"To me every hour of the light and dark is a miracle,
Every cubic inch of space is a miracle"*

Walt Whitman

## ABSTRACT

*This paper asks if a virtual space-time could appear to those within it as our space-time does to us. A processing grid network is proposed to underlie not just matter and energy, but also space and time. The suggested "screen" for our familiar three dimensional world is a hyper-sphere surface simulated by a grid network. Light and matter then travel, or are transmitted, in the "directions" of the grid architecture. The processing sequences of grid nodes create time, as the static states of movies run together emulate events. Yet here what exists are not the static states, but the dynamic processing between them. Quantum collapse is the irreversible event that gives time its direction. In this model, empty space is null processing, directions are node links, time is processing cycles, light is a processing wave, objects are wave tangles and energy is the processing transfer rate. It describes a world where empty space is not empty, space warps, time dilates, and everything began when this virtual universe "booted up".*

## INTRODUCTION

Previous work proposed that since the simulation hypothesis is about the knowable world, its contrast with the objective reality hypothesis is testable, based simply on which better explains our world [1]. This paper develops a design to simulate a space-time like ours, while later papers address light, matter and gravity, respectively

**Method**

The simulation conjecture contradicts the positivist supposition that nothing exists beyond the physical, but it doesn't contradict science. Only *assuming* that it is impossible short circuits science, which should evaluate hypotheses, not presume them wrong. That the physical world is a simulation seems absurd, but by thinking the unthinkable science has advanced in the past.

Driving this proposal is that objective reality assumptions no longer work in modern physics. In an objective reality, time doesn't dilate, space doesn't bend, objects don't teleport and universes don't pop into existence from nowhere. No-one would doubt that the world was objective if only it would act so, but it doesn't. Terms like "spooky" and "weird" well describe the quantum and relativity theories, that predict the world brilliantly but contradict our common sense ideas of it. Modern physics tells us that there is something rather odd about this world of ours.

Science "proves" a hypothesis by assuming it so, *following its logic*, then testing its predictions against world data. So the way to test this theory is to do just that, to *assume* it is true, *design* a hypothetical simulation, then *validate* it against what the world actually does. If successful, the core laws of physics will derive in a sensible way from information theory, illustrating Tegmark's "*Physics from scratch*" approach [2 p6]. The method is:

1. *Specify*: List the requirements of a world like ours.
2. *Design*: Design a feasible model.

---

3. *Validate*: Is the model compatible with the physical world?

4. *Repeat:* Until validation failure, logical inconsistency or further design is impossible.

The consistency constraint is significant, as a design can easily emulate *one* requirement but to emulate many at once is not so easy. The design should also:

1. *Follow best practices.* Use established computer science principles.

2. *Satisfy Occam's razor.* Given a design choice, the simpler option should be taken.

The research questions proposed are:

1. *Could a simulation emulate the physical world as we see it at all?*

2. *If so, is physical reality more likely to be a simulated or objective?*

While the inhabitants of a simulated world's can't see the programs creating they, they can see their world. If that world arises from processing, its nature should reflect that, e.g. it should be comprised of discrete pixels that refresh at a finite rate. Beings in a simulated world could look for the tell-tale signs of information processing. As we know how processing behaves and how the world behaves, this theory is open to evaluation by our science.

**Background**

A hundred years of research have validated quantum and relativity theories in sub-atomic and cosmic domains, yet they conflict at the core. The quandary is that:

1. *Quantum theory* assumes an objective space background, which relativity specifically denies. For quantum theory to satisfy relativity it must be *background independent,* i.e. not assume, as it currently does, that quantum states arise in a fixed space and evolve in a fixed time [3].

2. *Relativity* assumes objects exist locally, which quantum theory specifically denies. For relativity to satisfy quantum theory it must be *foreground independent,* i.e. not assume, as it currently does, that localized objects move relatively through space-time.

These two great theories contradict because each debunks an objective reality assumption the other still clings to. Quantum theory challenges the objective reality of foreground objects, but still assumes a fixed background. Relativity theory challenges space and time as objective backgrounds, but still assumes fixed foreground objects. Both theories rebelled against the idea of objective reality in different ways, so each exposes the other's conceptual baggage but ignores its own.

To reconcile, both theories must abandon entirely all objective reality assumptions, i.e. reject objective space, objective time, objective existence, objective movement and any similar ideas[2]. The prime axiom here is that nothing in the physical world exists of or by itself, so while it seems substantial and self-sustaining, *both its foreground and background arise from processing.* One can't "half-adopt" this theory, so it has no fixed space or time that quantum states exist in, nor any fixed entities that move relatively. The only constant is information processing, with space, time, matter and energy just outputs. In this model, not only is matter-energy calculated by "space" as Zuse suggests [4], but space itself is also calculated.

**The grid**

In our computer simulations, *programs* direct *a central processing unit* (CPU) to create the *pixels* we *observe*. In this model, quantum equation *programs* direct a *distributed processing network* called the grid to calculate quantum state *pixels* that are physical reality when *observed*. Yet there are two critical differences. First, the grid "screen", in our terms, that receives the processing is the same processing network that outputs it.

---

[2] This doesn't deny realism. The model doesn't say that the world isn't real, just that it isn't objectively real.

Second, the observer is not outside the simulation, as when we watch a video, but inside it, i.e. this system is simulating itself to itself. Figure 1 shows:

1. *Programs*. Entity class programs run *instances*[3] across the grid.
2. *Processing*. Grid nodes process them and pass them on, where:
    a. *Space*. Is the grid architecture.
    b. *Time*. Is the node processing sequence
3. *Pixels*. Quantum state interactions that overload the grid cause a node to "reboot", where:

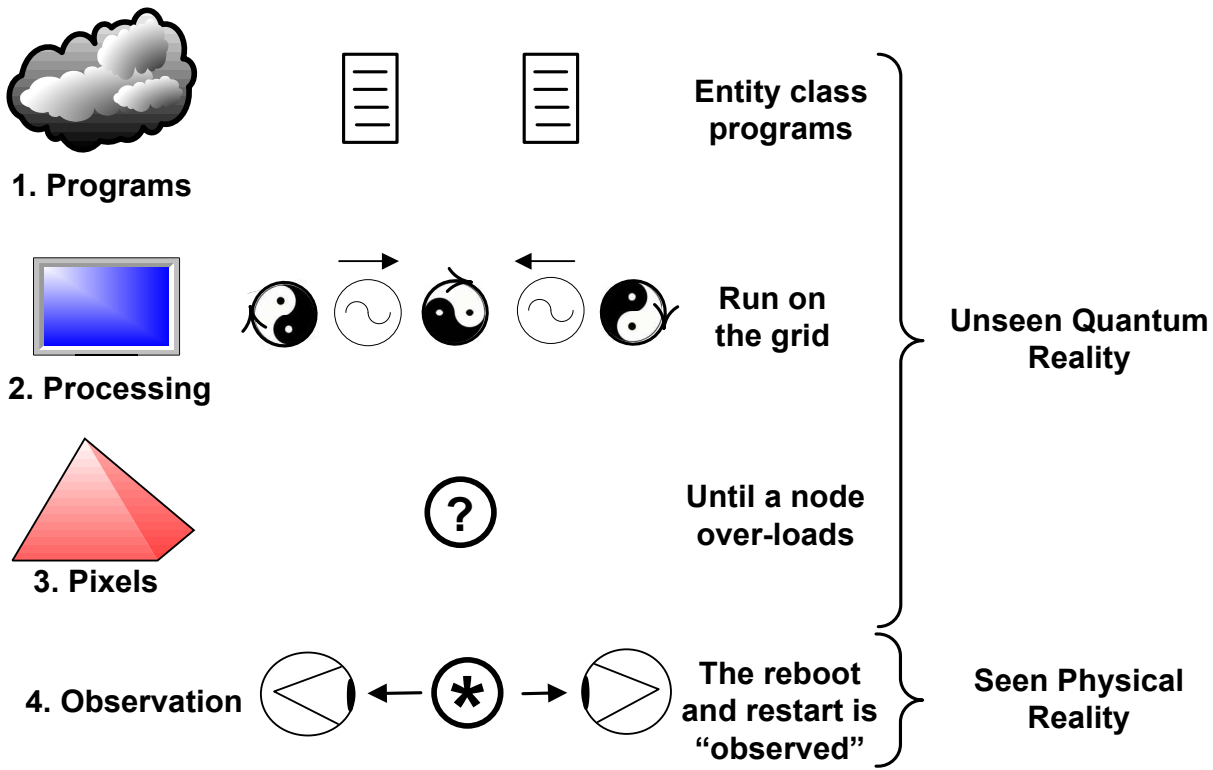

*Figure 1. A simulated reality model*

   a. *Light*. Is a transverse circular rotation moving in space.
   b. *Matter*. Is Planck energy light tangles in a grid node.
   c. *Energy*. Is the local grid node processing rate.
4. *Observation*. The local grid node reboot that restarts and re-allocates the processing of the entity classes involved is the event we call observing physical reality.

This paper considers how the structure and processing of the proposed grid explains strange features of our time and space and the next paper considers how observation as a grid node reboot explains the often "spooky" behavior of light. Yet the basic idea, that processing could create physicality, is not new:

---

[3] *Instantiation* is an object orientated systems (OOS) method where many information object "instances" inherit code from a single *class* source, e.g. screen buttons instantiating the same class look the same because they run the same code.

*Simulating space and time, Brian Whitworth*

1. *Fredkin.* His theory "*…only requires one far-fetched assumption: there is this place, Other, that hosts the engine that "runs" the physics.*" [5] p275.
2. *Wilczek.* Postulates "*the Grid, that ur-stuff that underlies physical reality*" [6 p111].
3. *Wheeler.* His phrase "*It from Bit*" implies that at a deep level, all is information.
4. *D'Espagnat.* Suggests a "*veiled reality*" beyond time, space, matter and energy [7].
5. *Tegmark.* His External Reality Hypothesis is that "*There exists an external physical reality completely independent of us humans*" [2].
5. *Campbell.* Proposes that "*The big computer (TBC)*" runs everything [8].
6. *Barbour.* Visualizes quantum waves as arising from an underlying landscape, where "*The mists come and go, changing constantly over a landscape that itself never changes*" [9] p230.

The *processing grid* of Figure 1 could be Fredkin's "other", Wilczek's Grid and Wheeler's "bit" (processing) from which "it" (the physical world) is derived. It fits D'Espagnat's view of a veiled reality behind the physical world, as when viewing a computer game the underlying screen is indeed "veiled" by the image upon it. The grid is also an external reality that computes the physical world, as Tegmark hypothesizes, it is Campbell's big computer, and Barbour's quantum mists could be patterns on a grid processing landscape.

As a city draws electricity from an power grid, so the physical world could draw its existence from a processing grid network. If so, *reverse engineering the grid design* then raises questions like:

1. *Architecture.* How do the nodes connect?
2. *Processing* What values do they set?
3. *Protocol.* How do they transfer network packets?
4. *Synchronization:* How do the nodes synchronize transfers?

## CONCEPTS

**Dynamic information**

What is information?

*Definition*

Shannon and Weaver began modern information theory by defining information as the power of the number of options in a choice[4] [10]. So a communication channel's bandwidth depends on the number of choices available, with two choices being one bit, 256 choices 8 bits or a byte, and one choice, which is no choice at all, is zero information. Information processing is then in turn defined as the transforming of stored information, i.e. setting a new choice value.

Now while a book is generally taken to contain information, its text is physically fixed the way it is. In itself the book exists as *one* physical choice, which by the above definition is *zero* information. At first this seems a false conclusion, but hieroglyphics that one can't decipher do indeed have zero information. A book only gives information if it can be read, when the reader's processing choices create information, e.g. the first symbol could be any alphabet letter, etc. A book's information *depends entirely on the receiver decoding process.*

So reading every 10th letter of a book, as in a code, gives both a different message and a different amount of information. If the encoding process is unknown the information is undefined, e.g. while the genetic alphabet is known the genetic language is not, as we are still learning how gene "words" enhance or suppress each other. To illustrate how information assumes decoding, consider a single electronic pulse sent down a wire. Is it one bit of information? If it means ASCII value "1" then it delivers one byte of information, or if it means the first word in a dictionary, say Aardvark, it gives many bytes, i.e. the information in a physical message per se is undefined. This explains how data compression can take the information stored a physical message and put it in a physically smaller signal - it uses more efficient encoding. Only if a reader uses the encoding as a writer do they agree on the information a message contains. Hence static information, as in a book, only exists in a dynamic processing context.

---
[4] Information $I = Log_2(N)$ for N options in a choice.

*Operation*

Let *static information* derive from an assumed decoding process, while *dynamic information* arises from the making of choices itself. So writing a book creates dynamic information, as one could write it in many ways, as does reading a book, as it can be read in many ways, but the book itself has no dynamic information, as it is just one way and no other. It does however have static information, but this only exists in the context of the dynamic process of reading it. In contrast, dynamic information is context free, and needs no external "reader" to define it. This distinction answers McCabe's argument that the physical world can't be simulated:

*All our digital simulations need an interpretive context to define what represents what. All these contexts derive from the physical world. Hence the physical world cannot also be the output of such a simulation* [11].

The logic is correct for our static information, which always needs a viewer context. However it is not true for dynamic information, as it doesn't need an interpretive context. So McCabe is right in that static information can't underlie our universe, e.g. imagine it "frozen" into a static state at a moment in time. It might then have static information, as a book does, but who would "read" it? Not us, as we would be frozen too. Without a reader, a frozen world is empty of information. Static information, like the letters on this page, is "dead" without a reader. Einstein discovered special relativity by imagining he was "surfing" a light wave "frozen" in space and time. Finding this was impossible, he changed our ideas of space and time instead.

*Implications*

Our physical world can't be a static simulation but it could be a dynamic one. *In this theory, the physical world is dynamic information not static information.* This conclusion, that the world we see arises from the making of choices not the storing of choices, has implications, e.g. that no quantum event is ever or can ever be permanently stored. A universe of dynamic information only exists by continually making choices, so like a TV screen image, disappears if not refreshed. The laws of physics fit this theory, as by relativity light can never be frozen and by quantum theory all entities quiver with quantum uncertainty. The model implies a world of continuous bubbling flux, which is exactly what we see. The only thing that doesn't change in our world is change itself.

However if such a simulation "leaks" choices, it will eventually run down, which our universe hasn't done over billions of years. This longevity requires one fundamental rule: *that the number of dynamic choices being made remains constant,* so the number of choices being made after a quantum interaction must the same as before. Indeed, a feature of our reality is its conservation laws, of matter, charge, energy and momentum, to which quantum theory adds spin, isospin, quark flavor and color. That only one law of conservation of processing exists is possible because all these conservations are partial, e.g. matter is not conserved in nuclear reactions and quark flavor is not conserved in weak interactions. In this theory, *the* law of conservation is *that dynamic information is always conserved*[5].

## The pixels of space

The *continuum problem* has plagued physics since Zeno's paradoxes over two thousand years ago [12]:

1. If a tortoise running from a hare sequentially occupies infinite points of space, how can the hare catch it? Every time it gets to where the tortoise was, the tortoise has moved a little further on.

---

[5] Except for the initial event, but see later.

2. Or, if space-time is not infinitely divisible, there must be an instant when the arrow from a bow is in a fixed unmoving position. If so, how can many such instants beget movement?

To deny the first exposes one to the second paradox, and vice-versa. The problem remains today, as infinities in physics equations that assume continuity, e.g. infinitely close charges should experience an infinite force and light with zero mass should go infinitely fast[6]. Only in a simulated universe of irreducible pixels and indivisible ticks do these infinities disappear, like ghosts in the day. In a discrete world, that there is no infinitely small stops the infinitely large from occurring. It follows that continuity is a mathematical convenience rather than an empirical reality:

*"… although we habitually assume that there is a continuum of points of space and time this is just an assumption that is … convenient … There is no deep reason to believe that that space and time are continuous, rather than discrete…"* [13] p57

The definition of information as choice assumes a finite set to choose from. As finite choices can't give an infinite continuum, a digital world must be discrete. Continuously dividing a simulated space must give a minimum "pixel" which cannot be split, and continuously dividing a simulated time must give a minimum "tick" which cannot be paused. So does our reality work this way? If it is continuous it can't be simulated.

Investigations of the continuity of our world break down at the order of Planck length, much smaller than an atom, and Planck time, much shorter than a light cycle. To examine such short distances or times needs short wavelength light, which is high energy light, but putting too much energy into a small space gives a black hole, which screens information from us. If you probe the black hole with more energy it simply expands its horizon and reveals no more. So in particle physics no-one knows what occurs below the Planck length. Just as closely inspecting a TV screen reveals only dots and refresh cycles, so closely inspecting physical reality reveals only Planck limits. If the world appears on a screen, then physicists know its resolution and refresh rate[7], which limits no technology will ever breach.

**Life on a brane**

Every simulation has a context, a containing reality with at least one more dimension than it has, e.g. our simulations appear on two-dimensional computer, TV and movie screen surfaces. The extra dimension is used to express and transmit the screen information. If a simulation always arises on a "surface" of a containing reality, our three dimensional space must be a surface in a four dimensional bulk:

*"When it comes to the visible universe the situation could be subtle. The three-dimensional volume of space might be the surface area of a four dimensional volume"* [13] p180

In 1919 Kaluza found that writing Einstein's general relativity equations for four dimensions of space produced Maxwell's electro-magnetic equations automatically, uniting quantum theory and gravity. This breakthrough was ignored at the time, as in four objective dimensions gravity would vary as an inverse cube, not an inverse square, so our solar system for example would collapse. That our world is virtual was not considered. To "fit" an extra dimension to an objective world Klein made it curled up in a tiny circle, so entering it just returned you to where you began. String theory uses six such "compactified" dimensions to model gravity. Yet why would an objective reality have six extra dimensions curled up inside it that it doesn't need? In this *existence geocentrism,* our world *must* contain any extra dimensions of physics.

In contrast, in this model, our world would be contained by any extra dimension, which would be too large for us see, not too small to see. The idea is that we are on a *brane* of a higher-dimensional bulk:

---

[6] By Newton F = m.a where F is force, m is mass, and a is acceleration. Hence a=F/m, so a force acting on a mass of zero (m=0) should give infinite acceleration (by that formula).

[7] Planck length of $10^{-33}$ meter is the pixel resolution. Planck time gives $10^{43}$ times per second as the refresh rate.

*"Physicists have now returned to the idea that the three-dimensional world that surrounds us could be a three-dimensional slice of a higher dimensional world."* [14] p52

Randall and Sundrum's local gravity model faithfully reproduces relativistic gravity using an extra dimension of infinite length, that is "sequestered" from our space [17], which this model explains as follows. If space is a surface, and our existence waves upon it, the dimension orthogonal to it is denied to us as a water wave is restricted to a pond surface. Water waves travel a lake surface by vibrating surface molecules up and down. The dimension at right angles to the pool surface must be free - if the pool top is sealed in concrete, it cannot vibrate so no waves can travel its surface. A wave needs a degree of freedom orthogonal to its surface to express its amplitude. Conversely, that amplitude direction cannot be a travel direction, i.e. it cannot leave the surface it moves on. So if we are quantum probability waves on a three-dimensional space surface, our wave amplitude is *to us* an "imaginary" direction, into which we cannot go. We are sequestered to our space for the same reason than an onscreen avatar cannot leave a computer screen.

In this model, the extra dimension beyond space wraps around our world rather than curls up within it. Our space is then just a surface in a larger bulk, and all existence is vibrations on that surface.

**Distributed processing**

An objective space with absolute Cartesian coordinates needs:

1. *A zero point origin*: A fixed centre, i.e. a (0,0,0) point.
2. *A known size:* A known maximum size, to set the coordinate memory needed[8].

Yet the galaxies in our universe are all expanding equally away from each other, not from a centre, to an as yet undefined maximum size. A Cartesian simulation would need an allocated size from the start*, i.e. before the big bang,* to avoid a Y2K problem[9].

Cartesian coordinates work for small spaces but *don't scale well* for universes like ours that expand for billions of years. In a scalable network, as the system load increases so does the processing needed to handle it [15]. The Internet is scalable because new nodes[10] that increase demand also add more processing. If network supply rises in tandem with network demand, the system can grow indefinitely. This scalability requires shared control, which is why the Internet has no "control centre". Initially some saw this as a recipe for failure, but sharing control lets systems evolve. And while an infinity anywhere in a centralized system will "crash" it, distributed systems localize problems. If grid processing is distributed not centralized, then each node only has to work with its local neighbors:

*"I remember … Richard Feynman … saying that he thought of a point in space-time as being like a computer with an input and output connecting neighboring points."* [16] p138

As space expands, new nodes add both more space and more processing, so the "performance" of space doesn't change as it expands - a desirable feature for a "real" universe. Distributing processing gives each node a finite processing capacity to manage its finite region of space, as in the spin network theory of loop quantum gravity [17]. In our world, the finite limit to the amount of information a finite region of space can represent is a black hole. That black holes expand when objects (information) fall into them supports the idea that space has a finite processing capacity, as does cosmology*:*

*"…recent observations favor cosmological models in which there are fundamental upper bounds on both the information content and information processing rate."* [18] p13.

---

[8] For example, the point (2,9,8) in a 9 unit cube space must be stored as (002,009,008) for a 999 unit space

[9] Before 2000 older computers stored years as two digits to save memory, e.g. 1949 was stored as "49". The year 2000 gave the "Y2K" problem, as the coordinates were all used up already. The year after 1999 was "00", which had already been used (for 1900). A lot of money was spent fixing this problem.

[10] The nodes of the Internet network are Internet Service providers, or ISPs.

A black hole then represents the maximum any grid node can process.

## SPACE AS PROCESSING

Knowing when entity packets in a node network "collide" is not a trivial issue. Two design options are:

1. *Entities calculate the interactions.* Each entity compares its position to that of all others, to know when it collides. Yet for *each* quantum entity in our universe to compare its position relative to *every* other at every moment is an enormous computing task. Also, light photons don't collide as matter does, but just combine their waves. If so, the simulated space arises from the objects in it.

2. *Nodes calculate the interactions.* Each grid node just processes a point of "space" each cycle. Now calculating collisions is much simpler, as it is just when overload the same node in the same cycle. Also, photons can just combine and not overload the node. If so, space in the simulation will exist as something apart from the objects in it.

As a processing simulation clearly favors the second option, which option does our reality favor?

The question of whether in our world space exists or not, apart from the objects in it, has concerned the greatest minds of physics. Simply put: *if every object in the universe disappeared would space still be there*? Is space "something" or is it truly nothing? Newton saw space as an objective canvas upon which objects are painted, so it still existed even without objects. Liebniz found empty space as a substance with no properties unthinkable, so argued that space is just a deduction we make based on object relations. To support this view, he notes that a vast empty space has no "where" for an object to be placed, unless there are other objects around. Also, that we only measure "distance" relative to other distances, e.g. a meter was the length between two marks on a platinum-iridium bar held in Paris. For Liebniz objects moved with respect to each other and "space" was imaginary, so if the universe of matter disappeared there would be no space.

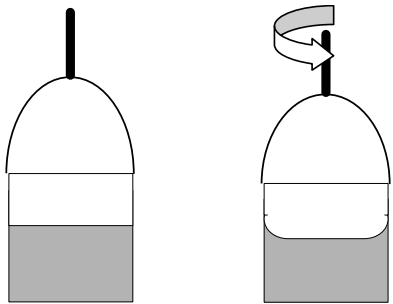

*Figure 3. Newton's' bucket.*

Newton's reply to Leibniz was a bucket - imagine a bucket filled with water hanging from a rope that is spun around (Figure 3). First the bucket spins but not the water, but soon the water also spins and presses up against the side to make a concave surface. If the spinning water moves with respect to another object, what is it? It can't be the bucket, as at the start when the bucket spins relative to the water the surface is flat. Only later when bucket and water spin at the same speed is it concave. In an otherwise empty universe where all movement is relative, Newton's spinning bucket should be indistinguishable from one that is still. Or consider a spinning ice skater in a stadium whose arms splay outwards due to the spin. One *could* see this as relative movement, as the stadium spinning around the skater, but if so the skater's arms would not splay. This suggests that the skater *really is spinning* in space [19] p32.

This seemed to settle the matter until Einstein upset Newton's idea of an absolute space through which objects move. Mach then tried to resurrect relative movement by arguing that the water in Newton's bucket rotated with respect to all the matter of the universe. According to Mach, in a truly empty universe the surface of Newton's spinning bucket would remain flat and a spinning skater's arms would not splay outwards! This reflects how unsettling to object orientated physicists is the idea that space, which one cannot touch or measure, is:

"… *something substantial enough to provide the ultimate absolute benchmark for motion.*" [19] p37

Rather than the objects creating space by their relations, as Leibniz thought, the current verdict is that "*space-time is a something*" [19] p75.

In this model, space, whether empty or full of matter, is the processing of an unseen grid. Yet it isn't the sort of objective backdrop Newton envisaged, as which node links to which pixel is not fixed. So an "object" that moves with respect to the objects around it need not move with respect to the grid, e.g. an onscreen avatar walking through an onscreen forest can stay at the same point on a screen because neither the virtual foreground nor the virtual background are fixed with respect to the screen nodes. Equally, a screen avatar could "see" other pixels by interacting with them, but not the screen that creates them.

**The architecture of space**

It seems strange to talk of the "architecture" of space, but computer simulations of it do just that:

"*…we think of empty spacetime as some immaterial substance, consisting of a very large number of minute, structureless pieces, and if we let these … interact with one another according to simple rules … they will spontaneously arrange themselves into a whole that in many ways looks like the observed universe.*" [20] p25.

This raises the strange question *What does space do?* Here is one specification:

1. *Existence*. Space allows objects to exist within it.
2. *Dimensions*. Space offers three degrees of movement freedom.
3. *Interaction*. Space defines when entities interact.
4. *Direction*. Space naturally moves objects in apparent straight lines (geodesics).

Can a transfer network meet these requirements, e.g. can the directions of space *derive* from grid architecture? The following does so, first for one, then two, then the three dimension case of our space.

*The Euclidean barrier*

Euclidean space is so deeply embedded in our minds that we tend to see it as the only way a space can be and structure all spaces accordingly, e.g. war-gamers divide their maps into hexagons that cover it completely, not octagons that don't. Loop quantum gravity [21], cellular automata [22] and lattice simulations [23] that use *static structures* to map nodes to points in an ideal Euclidean space cannot model a relativistic space that curves and bends. This needs a *dynamic structure,* that allocates nodes to points as the Internet allocates IP addresses - on demand.

The following figures seem show grid nodes in our space, but are just analogies to help understanding. To limit the grid that creates space by placing its nodes in the space it creates is backwards logic. We exist in space, but what creates space need not, e.g. our computers create web sites in cyber-space, not physical space, where "distance" is measured in mouse clicks not miles. Which node goes "where" is irrelevant to this architecture, as terms like "close" and "straight" arise from its connections, e.g. a node connected directly to another is "near" while one that is many links away is "far". If the directions and distances of our space derive entirely from how these nodes connect, they by definition exist outside the space they create.

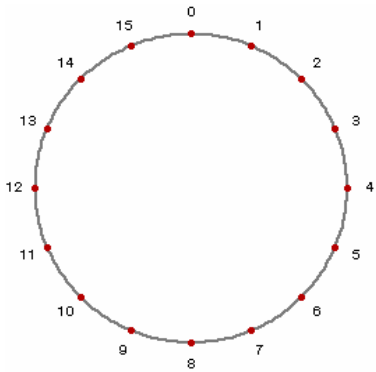

*Figure 4. One-dimensional space*

*One and two dimensions*

In Wilson's original idea of a lattice creating space, each vertice is a volume of space. From this Penrose developed the spin networks used in loop quantum gravity, where each vertice joins three event lines, so two inputs can give one output [24].

In the structure proposed here, the vertices are processing nodes and the lines between them information transfer channels, but while each node is still a "volume", it has many neighbors not just three.

If a set of processing nodes of equal capacity each connects itself to two arbitrary others, the result is a one dimensional space (Figure 4). Connecting one node to two neighbors gives two transfer directions, left and right. One notional rotation creates one dimension of space, as defined by the node connections.

To add another dimension, "rotate" the circle just created around an axis. Each node now forms another circle like the first, with the same number of points. This allows it two "orthogonal" movement dimensions, and the space can be visualized as a sphere surface with longitudes and latitudes (Figure 5). A two-dimension "flatlander" confined to points on the surface of this sphere would see a space that is:

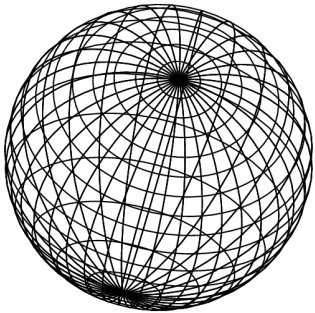

1. *Finite.* It has finite number of points.
2. *Unbounded.* Moving in any direction never ends.
3. *With no centre.* Each point is a centre of the sphere surface.
4. *Approximately flat.* If the sphere is large enough.
5. *Simply connected.* Any loop on it can shrink to a point.

*On demand connections*

How the Figure 5 nodes connect depends on the rotation axis used. A node on the rotation axis of the original one-dimensional circle (Figure 4) will become a *pole,* with many longitudes radiating out from it, each linking to a neighbor node. These nodes form a *planar circle* around the pole, and their connections define how it transfers information, i.e. its direction choices on the sphere surface.

*Figure 5. Two-dimensional space*

Yet the rotation making that node a pole was arbitrary. Rotating the original circle on a different axis gives the same sphere but a different connection configuration. The set of all possible rotations defines all possible connection configurations for the sphere surface. As all rotations involve the same nodes, changing the rotation axis only changes the node connections, which is easy for a network to do. Cell phone networks regularly change their architecture to improve efficiency under load.

Now suppose that *each node locally configures itself as a pole,* simply by setting its connections so. The coordinates of this space are defined on demand after a focal node is chosen, i.e. "just in time". There are no global grid node coordinates, as each node has its own longitudes and latitudes. Each node "paints" its own coordinates when activated, i.e. defines its own space. This doesn't allow an objective view, from without, but as this simulation is only ever seen from within, that doesn't matter. This apparent bug becomes a feature when relativity is considered.

As the connection requirements are symmetric, *every* node connects to a planar circle of neighbors around it, defining different directions out into the space. These directions, which are the node's transfer links, are veridical to an ideal sphere surface, i.e. correctly approximate it. So a two dimensional space can be modeled symmetrically by two consecutive rotations, if each node locally defines its own connections *as if it were the original rotation origin of the created space*.

*Network granularity*

The *network granularity* of the above space depends on the number of discrete steps in the rotations that create it. While a perfect circle has an infinite number of rotation steps, a discrete circle has only a finite number.

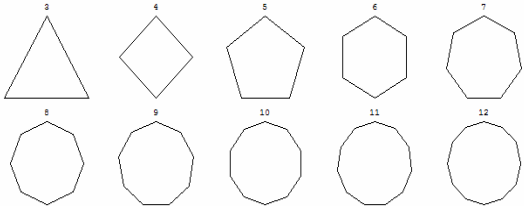

In this view, all polygons are discrete circles, e.g. a triangle is a "3-circle", a square is a "4-circle", a pentagon a "5-circle", and so on (Figure 6). So every polygon is an N-circle that increasingly approximates an ideal circle as N increases. Conversely, any discrete circle can decompose into triangles, so any discrete circle simulation can also be modeled by triangle-based simulations like spin networks.

Most space simulations use triangles to simplify the mathematics, but allowing only three movement directions is unrealistic. War gamers prefer hexagons to squares as they give six movement directions not four, but avoid octagons as they don't fully fill the board. What changes with granularity is the number of directions the space allows. Given no need for an objective Euclidean space, this model proposes a rotation of many steps to give many directions.

*Figure 6. Discrete rotations, N = 3-12*

Increasing network granularity improves its approximation of continuous space, but not all connections are reversible, so retracing a route taken back again need not return one to the same start node, though it will be a true vicinity. Also, even with the finest granularity, one can never fully cover a flat space with circles, so essentially this space has "holes" in it. Point objects, as electrons are thought to be, could pass right through each other. This would be a problem for the model if quantum entities existed at exact locations, but luckily in our world they don't. They exist probabilistically over an area, as smears not points, and so don't need exact locations to interact. Objects will register as returning to the same point if they return to the correct vicinity. *That quantum objects exist inexactly avoids the problems of an inexact non-Euclidean space.*

*Space as a hyper-surface*

The mathematician Riemann first wondered if our space was the three-dimensional surface of a *hyper-sphere*. He asked, if a mathematical hyper-surface is a three dimensional space, how can we say for sure that our space is not so? Are we, like Mr. A. Square of Flatland [25], unable to imagine a dimension beyond those we experience? If rotating a circle into a sphere generates two dimensions from one, rotating a sphere must, by the same logic, give three dimensions from two. We can't imagine "rotating" a sphere, but mathematically it is well defined. The result is a *hyper-sphere*, whose *hyper-surface* is an unbounded, simply connected three-dimensional space, just as our space is.

In this model the architecture of space arises from discrete rotation upon discrete rotation, giving an unbounded space that, like the surface of a sphere, has no edges or centre, and that expands equally over every point of its extent. The parallels with our space are strong.

The model predicts that the grid granularity, the number of connections per node, limits the number of directions of the space. That each node has a fixed number of connections predicts a minimum *Planck event angle* for single node quantum events[11], i.e. that direction, like length, is quantized.

---

[11] If a node has N neighbors in a circle around it, the minimum Planck event angle is 360°/N.

*Simulating space and time, Brian Whitworth*

If space is a hyper-bubble surface, it cannot be perfectly flat, as an absolute Euclidean space would be. In our world, general relativity clearly states that space is locally curved by gravity, which *requires another dimension for it to curve into*. If relativity is true and space curves, then it must itself be just a surface.

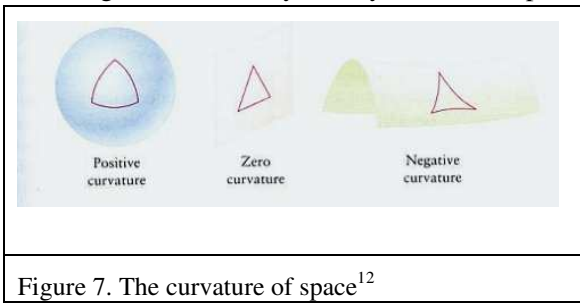

Figure 7. The curvature of space[12]

If space curves locally, cosmologists wonder if it is curved globally? One test is how a triangle's angles add up. In flat space they add up to 180°, with *positive curvature* they are more than 180°, and with *negative curvature* they are less (Figure 7). So the angles of a triangle laid out on the earth add up to more than 180°, as the earth's surface is positively curved. Of course for thousands of years we didn't detect that, so likewise the curvature of a universe that has expanded at the speed of light for billions of years may no longer be discernible. It is as near to flat as we can measure, but that is just the universe we see.

**Simulating existence**

In this model, quantum existence arises from processing waves. Imagine a pond with waves moving across its surface. The water molecules themselves don't move - they just vibrate up and down setting a positive and negative amplitude with respect to the surface. What moves across the surface is the wave *pattern*, which is information. Our "solid" world could arise at the quantum level in the same way – as a moving processing wave. Space as a hyper-surface allows the setting of positive and negative values, like bumps and dimples on a ball surface. Photon waves could then travel the three-dimensional surface of space as water waves travel a pond surface - by vibrating at right angles to it. Here the "medium" is a processing network, and the wave "amplitude" just arbitrary information values set.

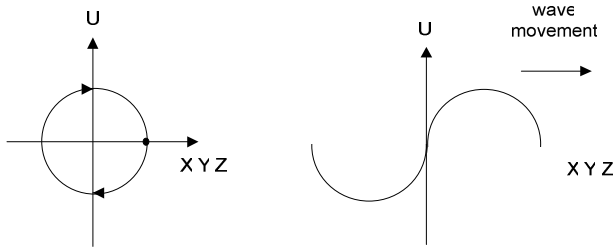

*Figure 8. Processing a. Space and b. Light*

Yet if the photons of our world are oscillations orthogonal to space, how can we ever detect them? We can if we exist as they do. If *all* existence is vibrations orthogonal to space, the world could interact as waves on a surface interact.

If space is based on a rotation, it makes sense to base existence in time on the same. The basic process proposed is a rotation of values transverse to space, i.e. a *transverse circle* (Figure 8a) could represent empty space as it is simple to calculate, gives a net surface displacement of zero and is continuous. In later papers, light arises when this function distributes across two or more grid nodes (Figure 8b), and matter arises when high energy light "tangles".

In this model, everything exists by oscillating in a dimension beyond space. In Schrödinger's equation, the core of quantum mechanics, matter is a three-dimensional wave whose value at any point is "something" the mathematic doesn't define, except to say that it "exists" in an imaginary complex dimension, i.e. outside our space. Schrödinger called it a "matter density" wave, because high values mean that matter is more likely to be there. Born called it a probability wave as its amplitude squared is the probability the entity exists at that point. One would expect the ultimate formula of an objective reality to represent something physical, but it doesn't. It is in itself just information, a number that doesn't derive from mass, momentum, velocity or any other physical property, yet the physical properties of the world derive from it. A physical world that emerges from quantum probabilities supports the idea that *substantial matter arises from insubstantial information*.

---

[12] From http://universe-review.ca/index.htm

*Simulating space and time, Brian Whitworth*

**A transfer algorithm**

In our world an unforced object travels in a straight line, defined as the shortest distance between two points. The general term is a *geodesic*, e.g. on a curved surface like the earth the shortest distances between points are longitudes and latitudes, which are geodesics even though they are curved. Geodesics are the lines along which objects naturally move. They define space, so changing them changes space, e.g. gravity "curves space" by changing the geodesics. A "point" in space is defined by the geodesics passing through it at an instant of time:

"*A point in spacetime is then represented by the set of light rays that passes through it.*" [26] p110

In this model, the directions of space arise from how grid nodes connect. Each node is linked to neighbors, which links *approximate* the directions around it. How these nodes receive and pass on entity information packets defines the constant passage-ways of space we call straight lines. A transfer rule is now proposed, first for two dimensions then for three.

*Two dimensions*

Any distributed communications network needs transfer rules, that tells grid nodes how to pass on packets. When a node receives a processing packet from a neighbor, which other neighbor does it pass it on to? What defines this defines the geodesics of the space.

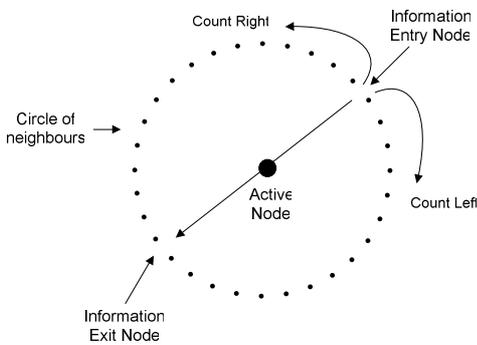

*Figure 9. Planar circle entry/exit nodes.*

Previously, for two dimensions, each node made itself a rotation "pole" by linking directly to a local *planar circle* of neighbors (Figure 5). So the problem is to find the "out" node for any planar circle "in" node.

If each node transfers input to the *opposite* planar circle node, as in Figure 9, any set of such transfers will be the fewest possible for that route, i.e. be a "straight line". Maximally separating entry and exit nodes on the planar circle minimizes the grid transfers for a route. A node can do this by counting both ways from the entry point until an overlap occurs, which is the exit point. This simple algorithm gives straight lines in a two dimensional space.

*Three dimensions*

In three dimensions each node has a sphere of neighbors not a circle, but suppose that *all transfers still occur in planar circles*. This is not unthinkable, as quantum Hall models use two-dimensional excitations called *anyons* to derive quantum events [27]. If so, a photon's polarization plane is then be its planar circle transfer channel. If all quantum transfers use planar circle channels, the transfer problem in three-dimensions becomes how to keep transmissions in the same plane. As one can cut sphere into many planes through one surface point, so a node receiving input would have to decide which of its many planar circle channels to use to generate the output. This, as before, is essentially an in/out problem:

G*iven a sending node planar circle, which receiving node planar circle should process it?*

A simple rule is that the receiving planar circle must overlap points of the sending planar circle and the sending node itself. That only one planar circle does this derives from the connection architecture. In spatial terms only neighbor circles in the same plane overlap three points (Figure 10). If a receiving node can use the incoming planar circle to pick one of its planar circles process the packet, a chain of planar circles can link together into a straight line movement channel (Figure 10).

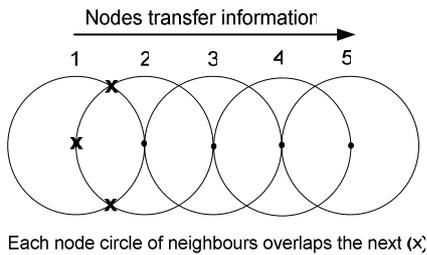

*Figure 10. A planar channel*

*Summary*

A grid architecture that links overlapping planar circles, plus a transfer algorithm that maximally separates entry/exit nodes, can simulate straight lines in space. The processing load differential of gravity will later be argued to bend light by skewing this transfer. Distinguish the *transverse circles* outside space (Figure 8) from the *planar circles* inside space (Figure 9). Both are discrete grid circles, but the first processes existence in time while the second processes movement in space.

## TIME AS PROCESSING

Does an extra dimension plus four dimensions of space-time give five dimensions in all? In this model, the dimensions of time and existence are one and the same, so it has only four degrees of freedom, three for space and one for existence in time. It supports the Hartle-Hawkin no-boundary theory that the big bang began when one of *four equivalent dimensions* "somehow" became time while the other three became space [28]. That "somehow" is here proposed to be by becoming the dimension that processes the cycles of existence that create our time.

**Simulating time**

Objective time should pass inevitably by its own nature, not needing anything else, while simulated time always depends on the processing, e.g. in John Conway's "Life" simulation (Figure 11) pixels reproduce and die by program rules, so blobs grow and contract until (often) a steady state is reached[13]. For a pixel entity within the Life simulation, time is measured by the events that occur to it, i.e. many events constitute (for it) a long time, while a few events are a short time. We measure time like this in our world, as atomic clocks effectively count atomic events.

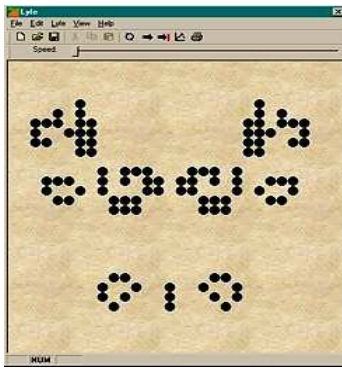

*Figure 11. A Life simulation*

Suppose a Life game that usually takes twenty minutes to reach a certain state is run again on a faster computer and reaches the same state in only two seconds. Running the game again takes less time in our reality but the passage of simulated time does not change, as the same number of events occurred. A pixel avatar in the simulation, seeing the same number of events passing, will conclude the same amount of time has passed. The passage of time in a simulation doesn't depend on the passage of time in its containing reality. *Simulated time depends solely on the number of processing cycles that occur.*

If a computer screen slows down due to processing load, an external observer sees it slow down, but an onscreen avatar sees no difference, as they themselves also slow down. So if our world is a simulated reality, we can't in principle see load effects, and indeed relativistic changes in space-time are undetectable to the parties affected. However if the screen processing is distributed, as proposed here, those affected can compare notes, to see that time differences really did occur.

For example in Einstein's twin paradox, a twin travels the universe in a rocket that accelerates to near the speed of light, and returns a year later to find his brother an old man of eighty. Neither twin was aware their time ran differently and both still got their allocated number of life breaths. Yet one twin's life is nearly done while the other's is just beginning. In particle accelerator experiments time really does slow with speed, so this paradox is not fantasy.

---

[13] See http://abc.net.au/science/holo/lablife.htm.

In this model, grid nodes have a finite bandwidth, so given another processing task, like movement, they process existence in time more slowly. As the rocket twin's rapid movement loaded the grid, it could only manage to process one year's worth of events for him. He just sees a normal year of events pass by, unaware that his time is dilated. In contrast, the grid processing his twin on earth has no such load, so eighty years of his life events cycle by in the usual way. Only when the two re-unite is it apparent that their time, which is processing, ran at different rates. Only in a simulated reality could this occur.

A specification to simulate a time like ours could be:

1. *Sequential.* Time creates a sequence of events.
2. *Causal.* Time allows one event to cause another.
3. *Unpredictable.* Time sequences are unpredictable.
4. *Irreversible.* Time can't go backwards, except for anti-matter.

The simulated time must be sequential, causal, unpredictable and irreversible.

*Sequential*

An objective world could digitize time as a sequence of existence states, as movie sequences of static pictures run together seem life-like. Time passing would then be a finite sequence of objective states that exist, as in Zeno's idea of movement as a sequence of indivisible object instants. The calculus of Leibniz and Newton could then work because things really do vary in space or time infinitesimals, i.e. one could replace all the delta time or space values in physics equations with delta-sequence values[14].

Time as a discrete sequence of objective existence states leads Barbour to conclude, logically, that time is itself timeless, being a "time capsule" of instances, that like the pages of a book can be turned back and forth [9] p31. Yet if past, present and future states exist timelessly, is life then a movie already made? If so, why is the system bothering to run it? Or as argued earlier, who could read a timeless capsule of past, present and future? Certainly not us, who are pictures on its pages. That time is a sequence of objective states stored in a permanent database strikes two problems. First, the number of states of a universe of quantum interactions are innumerable. Any simulation that stores its past in a static database soon meets storage problems. Secondly, what would be the point of storing every *quantum event* in a vast database? Even if we could view this "record", it would be like getting to know someone by reading the three billion letters of their DNA, or viewing World War II as a series of atomic events. That the system somehow sorts out what is "important" to save is an equally intractable problem, as by the butterfly effect of chaos theory, a single photon could change all history[15].

*Causal*

In contrast, this model has no static states to exist timelessly or otherwise, and uses no static storage at all. The only "permanent" record of the past is the dynamic present. Hence, as our brains recall the past *now*, so the "memory" of fossil rocks records *today* what occurred when dinosaurs roamed the earth, and before. Equally, our DNA reflects not just our ancestors, but the entire human race, if not all life on earth. In this evolutionary storage system, genes [29], memes [30] and norms [31] persist for thousands or millions of years by their generative power, while that which lives only for itself fades away. If what exists now *is* a record of the past, the database that stores it is the present. The system decides what worth keeping in the record by its ongoing choices.

In this model, the sequence that creates time is not of states, but of processing operations that, like the statements of a computer program, set the values we call states. This processing gives causality, as in quantum

---

[14] Replace each *dt* or *dx* term in calculus with a *ds* term, a small extent in a finite sequence of states. Now as *ds*, the number of intervening states, "tends to zero" it can become zero when there are no more intervening states.

[15] A human eye can detect a single photon, and one person can change the world, so a photon could change the world, i.e. every photon is potentially "important". How could one know which one actually is?

*Simulating space and time, Brian Whitworth*mechanics a state can: "*… evolve to a finite number of possible successor states*" [32] p1. If time is a processing sequence, not a state sequence:

*"Past, present, and future are not properties of four-dimensional space-time but notions describing how individual IGUSs {information gathering and utilizing systems} process information."* [33] p101

However processing, like the formulae of quantum mechanics, is entirely mechanical.

*Unpredictable*

Yet if the quantum world is a machine, inexorably grinding every option, it is a machine with:

"*…roulettes for wheels and dice for gears.*" [34] p87

Knowing everything physically knowable, physicists cannot predict when a radio-active atom will emit a photon. It is *random,* i.e. indeterminate with respect to it's physical past. No story of preceding physical events gives rise to it. So repeatedly querying a entity's quantum state, like its spin, gives an unpredictable physical answer each time. If *free* means unconstrained by the causal physical world, and if *choice* means taking one of two or more options available, a random event is a *free choice.* So while quantum mechanics describes a machine, quantum collapse implies free choice, which in turn implies information.

A choice that creates information, by definition, has a "before" and "after" - before the choice there were many options, but after there is only the chosen one. The link is not only causal, but also unpredictable. If it is predictable, there is just an inevitable sequence, with no choices. Before a choice, the outcome is by definition unknown, as if it is already known, there is no choice at all, i.e. zero information. That there is information at all requires choice, which quantum collapse provides. This denies a "timeless" future set in place. The world machine can only be predicted probabilistically because the choices that produce it are uncertain until made. Time as a sequence of static states is causal but predictable, and as static information, needs a decoding context. Time as processing punctuated by interaction choices is causal and unpredictable, and as dynamic information, it is context free. It allows neither past nor present, but only a "*Physics of Now"* [33], where each so-called static state is just another choice.

*Irreversible*

As the laws of classical physics are all reversible in time, physicists have never been clear why time itself can't run backwards. Time as sequence of states run forwards should also be able to run backwards. The same is true for time as processing, unless it has an irreversible step. The collapse of the quantum wave function is a one-way, irreversible step, so any event sequence containing it, like any observation, is also irreversible. So by this model, one can neither run time forward to prophesize a predestined future nor run it backwards to revisit the past. Einstein's proof that our universe has no common time does not make any time possible. That grid nodes cycle at different rates lets time go slower or faster, but doesn't alter the *processing sequence.* Causality and unpredictability are preserved, as the following paradoxes show:

1. In the *grandfather paradox* a man travels back in time to kill his grandfather, so he could not be borne, so he could not kill him. One can have causality or travel back in time, but not both.

2. In the *marmite paradox* I see forward in time that I will have marmite on toast for breakfast, but then choose not to, so didn't rightly see forward in time. One can have choice or predictability but not both.

Time as processing punctuated by random choices is sequential, causal, unpredictable and irreversible, i.e. does not allow time travel.

*Anti-matter*

That all matter processes time in the same direction is a reasonable condition for a stable space in computer simulations [20]. Yet if time arises from the processing steps of a fixed transverse circle, these values aren't irreversible, so a single entity, *between interactions,* could run its existence processing in reverse. In a later paper, normal matter exists by clockwise processing but anti-matter uses anti-clockwise processing to exist. In Feynman diagrams, anti-matter enters events going backwards in time. In this model, that is because it exists by

*Simulating space and time, Brian Whitworth*

running the processing that creates our existence in reverse. There is no absolute time, so while anti-matter runs *our* virtual time in reverse, it can no more undo its interactions than matter can. Anti-matter goes backwards in our time, not its time, so to an anti-matter entity it is our time that runs backwards[16]. Time as processing clarifies that when anti-matter "runs time backwards" it just exists by running our processing sequence in the opposite way.

This also explains Einstein's statement that for light, time itself stands still. If nodes "tick" from one state to the next, *no time event can occur in less than a node cycle*. Our time then is the number of processing cycles a grid node completes. If a photon moves at one node per tick, it goes to the next node before an event can occur. It never experiences a node cycle so time never passes for it, i.e. its time stops.

### Summary

Time in our world isn't an objective flow carrying all before it, though superficially it seems so. That it is a processing output explains why it varies locally. Space as an objective backdrop to fixed objects isn't how our world works either, though again superficially it seems so. Space as grid network transfer channels again better explains how it "curves" and "stretches" under load. In this model, space and time do not exist objectively, but are the by-products of grid processing:

"*… many of today's leading physicists suspect that space and time, although pervasive, may not be truly fundamental.*" [19] p471.

Objective space and time are convenient concepts for ordinary life, but don't explain the extraordinary world of modern physics. Even today, when people first hear Einstein's theory that time and space are malleable, they suspect some sort of intellectual trick. Yet is no trick. It is not our *perceptions* of time or space that change but *actual time and space* as measured by instruments. What could possibly explain this, other than that our time and space are indeed simulated?

## IMPLICATIONS

That space and time are simulated has implications for physics.

### The big bubble

Astronomers see all the stars and galaxies receding away from us, so either we just happen to be the centre of the universe (again), or the view is the same from all vantage points. Yet how can space have its "centre" everywhere? In cosmology, the big bang created a space that has been expanding ever since, but how can a space "expand"? If the big bang exploded from a singular point in an objective space, why isn't the energy of this "explosion" at the edge of the universe by now? Why is the cosmic background radiation of the big bang, still everywhere around us today, visible as static on blank TV screens? The idea of objective space poorly explains big bang cosmology.

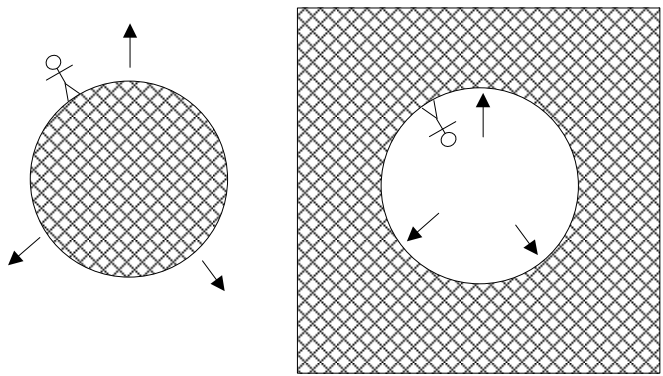

*Figure 12a. The Big Bang, 12b. The Big Bubble*

Space as a hyper-sphere surface fits the facts, as like a sphere surface it has no centre or edge. Travel in any direction, as on the earth's surface, eventually returns one to the same place. Such a space would also expand equally at every point, and an explosion on its surface would first go "out", then wrap

---

[16] The direction of processing of our matter was set by the first grid node that split in the big bang. This was the original "symmetry breaking". So there was no balance of matter and anti-matter that resolved in our favor.

*Simulating space and time, Brian Whitworth*

around to end up everywhere. Radiation from a nuclear explosion on earth would do this. So cosmic background radiation is still here because it has circled the universe, perhaps many times.

The term "Big Bang" places us on the *outside* of a spherical expansion (Figure 12a), but that we are *inside* an expanding bubble (Figure 12b) allows answers to questions like:

1. *What is space expanding into?* It is expanding into a surrounding four dimensional bulk.
2. *Where is space expanding?* Everywhere - new grid nodes fill "gaps" that arise everywhere.
3. *Where does the new space come from?* From the bulk around the bubble.
4. *Are we expanding too?* No, existing nodes are not affected when new nodes are added.
5. *Was the entire universe once at a point singularity?* No. It began as a unit sphere not a unit point.
6. *Why didn't the new universe immediately form a black hole?* A big crunch contracting the current universe would soon form a massive black hole, so why didn't the big bang immediately do the same? In this model, it didn't because it didn't begin all at once (see following).

*The first event*

Suppose, in the first event, a <u>single</u> grid node somehow "split apart", giving its processing to its neighbors, and leaving a "hole" in the bulk, across whose the inner surface it could "move"[17]. No black hole would occur, as the processing of one node can't overload all those around it, i.e. the event created more space than light. As this "first light" divided over more grid nodes, by the expansion of space, the electro-magnetic spectrum ensued (see the next paper). In the big bang <u>one grid node split to give the first light on the hyper-surface of space</u>.

This initial "rip" then spread, like a tear in a stretched fabric, in a cataclysmic chain reaction that respected no speed of light limit. It occurred at the node transfer rate not its cycle rate[18]. Physicists extrapolating the universe's expansion back in time find that it did indeed initially expand faster than the speed of light, a period called *inflation,* but currently have no reason for it [35]. In this model, *inflation was when the grid itself was ripping apart.*

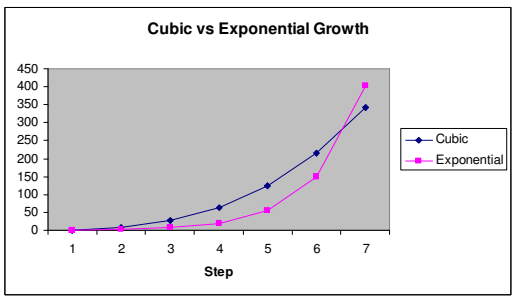

*Figure 13. Cubic vs. exponential growth.*

Equally problematic is why inflation stopped? Suppose the grid split apart in a classic exponential chain reaction, while the the hyper-surface of space expanded as a cubic function. Which growth predominates? If the chain reaction does the rip won't stop, but the expanding space can stop the rip as it dilutes waves on its surface by inserting points into their wavelength. Hence cosmic background radiation, white hot at the dawn of time, is now cold.

As Figure 13 shows, a cubic growth initially wins over an exponential one, but then is quickly overpowered. Yet the two functions are not independent - the growth of space dilutes the light causing the chain reaction. Even so, the resolution will be quick, and indeed by some cosmology estimates inflation was over in the first millionth of a second or less of creation. In So the expanding bubble (of space) weakened the waves on its surface enough to stop the chain reaction. The grid injury healed itself, but the hyper-bubble created continued to expand at light speed, simulating our universe on its inner surface.

---

[17] In this model, the inner surface of the hyper-bubble is our space, and the processing moving is light.

[18] In this model, grid nodes normally transfer once per processing cycle, i.e. at the speed of light. In this case, each transfer immediately splits the receiving host node apart, before it can process anything. The chain reaction is faster than any node processing cycle, i.e. faster than the speed of light.

*Simulating space and time, Brian Whitworth*

In this model, *all* the free information of the universe arose from a once only inflation chain reaction that will not repeat [29]. Since then galaxies formed and space expanded, but never again did the grid itself rip apart. So the free information of the universe is constant, as since inflation no more has been added. This avoids the illogic of a "real" universe that meticulously conserves itself coming into existence from "nothing", as big bang theory implies. It predicts that the total information of our universe is, and since inflation always has been, constant.

So to the principle of dynamic information conservation, given earlier, is added: "except during inflation". In current theory, matter converting to energy implies a deeper conservation beyond matter or energy. Equally, that time and space trade-off with matter-energy implies a quintessence of processing. What now is conserved is the grid, whether as the message of light or the unseen medium of space that delivers it.

**The synchronization problem**

In solving some problems, this model creates others, as a processing network must synchronize its transfers, e.g. if a sending node transmits two "photon events but a receiving node completes only one receive photon event, the second photon disappears. Dynamic information has no inherent reality, so what is not received ceases to exist. If computer simulations "lose" transfers, simulated objects vanish for no apparent reason - forever. We solve this problem in two ways:

1. *Centralization*. Synchronize transfers by central control.

2. *Buffers*. Allow asynchrony by giving each node a memory buffer.

*Centralization*

Central processing unit (CPU) chips illustrate the transfer synchrony problem. If a CPU issues a command to move data from memory into a register, how does it know *when* it is there? The CPU cycles much faster than the rest of the computer, so must wait many cycles for anything else to happen. If it reads the register too soon, it gets garbage left over from the past, but if it waits too long it wastes its own processing cycles. Can it check if the register data has arrived? Issuing another command needs another register for the results, which then also needs checking!

Our computers solve the problem by synchronizing all events to a central clock, like a conductor keeping an orchestra in common time. The chip just waits a fixed number of *clock cycles,* then *assumes* the task request is done, i.e. reads the register. Hence users can "over-clock" a computer by increasing its clock speed from the manufacturer's safety default. This avoids asynchrony by making every system element, fast or slow, run at the same clock rate. A universe simulated like this would run at the speed of its most overloaded node.

*Buffers*

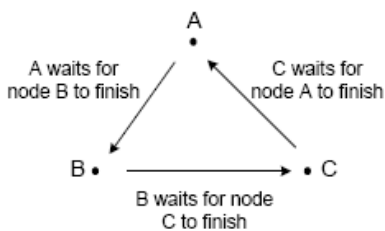

*Figure 14. Transfer deadlock.*

Networks that centralize control are reliable but slow. Modern networks, like Ethernet, run faster by giving nodes autonomy, e.g. on the Internet, each node run at its own pace. because if another node transmits when it is busy, it stores the input in its buffer memory.

In computing, a buffer is the memory a fast device uses to connect to a slower device, e.g. a computer (fast device) sends data to a printer (slow device) by a printer buffer, so you continue working while the document prints. So if an Internet node is busy, it holds the input in a buffer until it can process it.

However buffers only work if one chooses their size and location correctly. If they are too big, they waste memory for no value, while if too small, they overflow so transfers must wait. This allows transfer "deadlock", where A waits for B, which waits for C, which is waiting for A (Figure 14). If our universe worked this way, a part of space could, like a screen dead spot, become unusable forever. And while known Internet backbone servers allocate bigger buffers, where galaxies will be in our space is unknown. A simulation of our universe that used static buffers would have serious problems.

*Simulating space and time, Brian Whitworth*

*Pass it forward*

If centralization is too slow and buffers are unreliable, how can a dynamic, distributed grid handle synchronization issues? If a node transfer *waits* for a destination node to finish it's current cycle, the speed of light would vary for equivalent routes, which it doesn't. The alternative is that transfers *never wait,* i.e. a *pass-it-forward protocol,* when a node passes a processing cycle packet immediately to the next, which must receive it. But what if sender and receiver nodes are asynchronous? If a node immediately accepts any new processing sent to it, what happens to its current processing? Won't accepting new packets on demand "lose" its existing ones?

Not if the receiving node also passes its processing on to the next node, which can also do the same, and so on. This could create an infinite regress, except that space is expanding, i.e. adding new nodes. Any pass-it-forward ripple will stop if it meets a new node of space, that accepts the extra processing without passing anything on. In this protocol, nothing on the grid ever waits. Light always moves forward at one node per local tick. There are no static buffers, as every packet passed on is accepted. The expansion of space avoids infinite pass-it-on loops.

The light that fills space then helps synchronize sender and receiver grid nodes, but the cost is in increase in the number of node transfers relative to node processing cycles. Where interactions are common, as near matter like the earth, the effect is minimal. However for light traveling the vastness of space, these minute adjustments add up. The effect is that less processing runs, i.e. less energy. This links to the cosmology finding that 73% of the universe is an unknown "dark" energy, that arises somehow from the vastness of space (Ford, 2004, p246). *if dark energy arise from the asynchrony of space, it is just an artifact of the system, not an actual existence..*

**Empty space is full**

In an objective reality empty space should be empty of energy, but in quantum theory:

> "… *space, which has so much energy, is full rather than empty.*" [36] p242.

In this view, empty space is the processing power behind a black hole that just happens to be running a null cycle. Every grid node, whether running space or matter, has the same capacity. As an "idle" computer still actively "decides" billions of times a second to do nothing, so the apparently empty points of space run their null cycle (Figure 8a). If the grid never stops processing, empty space is not empty [37]. Consider some evidence:

1. *Virtual particles.* In quantum theory, space itself creates virtual particles and anti-particles. They "borrow" energy from the vacuum, exist briefly, then equally quickly disappear back into it. In the *Casimir effect,* two flat plates in a vacuum placed close together experience a force pushing them inwards. This "pressure" is attributed to virtual particles in the vacuum around the plates pushing inwards, which particles cant arise between the plates when they are closer than their wavelength. Null processing can generate virtual matter and anti-matter particles, because equal cycles of opposite rotation are also null.

2. *Vacuum energy.* What physicists call the energy of the vacuum arises in quantum theory because a point can't have a fixed energy, i.e. exactly zero energy. It can only average zero. A null processing cycle, of equal positive and negative values, likewise *averages* zero, but isn't a constant null value.

3. *The medium of light.* How can light, that physically exists, vibrate a vacuum that is "nothing"? If space itself is the medium of something, then it cannot itself be nothing. A vacuum that transmits no physical matter or energy may seem "nothing" to us, but if the potential to transmit remains, it cannot actually be nothing. If space is the null processing that mediates light, empty space is like a screen that is blank but still switched on.

Does this imply an "ether"? A physical substance permeating all space would give a standard reference frame to all movement, which relativity denies. The Michelson–Morley experiment denied the idea of a *physical* ether, but the idea of a *non-physical* ether has never been contradicted:

> "*Since 1905 when Einstein first did away with the luminiferous aether, the idea that space is filled with invisible substances has waged a vigorous comeback.*" [19] p76

*Simulating space and time, Brian Whitworth*

Indeed, even Einstein later concluded that without some sort of ether, relativity was unthinkable [38]. Likewise, many in quantum theory propose a "new-ether":

"*The ether, the mythical substance that nineteenth-century scientists believed filled the void, is a reality, according to quantum field theory*" [39] p370.

In this model, the "medium" of the physical universe is the grid that processes it.

*A perfect transmitter*

When we view empty space we see nothing, just emptiness. This could mean nothing is there, as objective reality supposes, or that a processing host perfectly transmits all light and matter, as proposed here. When one looks out of a transparent window, the glass transmits light from objects behind it. One sees the message the light transmits, but not the glass medium sending it. One only knows the glass is there if it has imperfections, by its frame surround, or by touching it.

Now imagine a world filled by a perfect transmitter with no imperfections so it can't be seen, that is all around us, so it has no boundary, and that even transmits matter, so it doesn't repel touch. If physics is processing, this is not impossible. If such a medium filled every direction, one couldn't see around it. If it passed on all light perfectly, it would itself be unseen. And if we moved into it, it would just pass on the matter of our bodies on as it does light. It would be like a *perfect diamond* completely filling all space, continuously and flawlessly reflecting the images of physical reality within it.

## CONCLUSIONS

Over a century ago, relativity and quantum theory left the safe haven of traditional physics for a conceptual wilderness of existence waves, quantum uncertainty, higher dimensions and malleable space-time. The journey produced amazing mathematical tools, but today fundamental physics wanders a semantic desert, filled with mathematical thorns. The advance of string theory, parched of empirical data, has ground to a halt. What began as a theory of everything has become a theory of nothing, going nowhere [3]. Some say it speculated too far, but here it didn't go far enough. It could not see past the conceptual mountain range of objective reality, but the "gaps" are there for all to see:

1. *Quantum randomness* is independent of the physical world, so perhaps is from outside it.
2. *Complex rotations* explain light, so perhaps photons do rotate in an imaginary dimension.
3. *Kaluza's* extra dimension unites relativity and Maxwell's equations, so perhaps it exists.
4. *Quantum waves* are probabilities, so perhaps the universe they produce is numbers too.
5. *Planck limits* on space and time suggest a discrete world, so perhaps it is.
6. *Calculus* assumes infinitesimals approximate reality, so perhaps they do.
7. *The uncertainty principle* denies fixed objective properties, so perhaps there aren't any.
8. *Feynman's* sum over histories assumes quantum particles take every path, so perhaps they do.
9. *Special relativity* lets time dilate, so perhaps it is processing slowing down.
10. *General relativity* lets space bend, so perhaps it is a transfer affected by a load differential.
11. *Cosmic background radiation* is still all around us, so perhaps space is a hyper-surface.

One is reminded of the duck principle:

*If it looks like a duck and quacks like a duck, then it is a duck.*

Why not take the equations of quantum theory and relativity as literally true? We use them to predict, so why deny what they imply? If the world looks like a simulation, by what scripture of science is it not? The reader can decide (Table 1) if our space or time contexts are simulated or objective.

*Simulating space and time, Brian Whitworth*

If they are, then instead of *things* we have *choices,* in a world where one can take but not hold, use but not keep and arrive but not permanently stay. If the flux of choice never stops, there are no ideal Utopian "end states - only the journey exists. Now there are no "states" at all, just a here and now choice, that one cannot "possess" but only make. In a dynamic processing world, there is only the *ever-present here and eternal now*.

*Table 1. Virtual and physical properties for space and time*

| **Virtual Property** | **Physical Property** |
|---|---|
| *Processing flux.* A processing simulation:<br>a) Continuously makes choices<br>b) Conserves dynamic information | *World flux.* The physical world:<br>a) Is always and everywhere a changing flux<br>b) Has many partial conservation laws |
| *Pixels.* In a simulation:<br>a) Nothing is continuous<br>b) Space arises from discrete grid nodes<br>c) Time arises from discrete processing cycles<br>d) Direction arises from discrete connections | *Quantization.* In our world:<br>a) Continuity creates paradoxes and infinities<br>b) Space is quantized at Planck length<br>c) Time is quantized at Planck time<br>d) Direction is quantized? (Planck angle) |
| *Null processing.* Null processing:<br>a) Produces no output<br>b) Is continuously actively running<br>c) Equates to equal and opposite cycles<br>d) Can host other processing | *Empty space.* Empty space:<br>a) Looks like nothing<br>b) Has vacuum energy<br>c) Spawns virtual particles<br>d) Is the medium of light and matter |
| *Distributed processing.* Allows:<br>a) A system to scale up well<br>b) Each node to paint its own connections<br>c) Each node to cycle at its own rate<br>d) A finite processing capacity for each node | *Localization.* Local space is consistent:<br>a) The laws of physics are universal<br>b) Each point has its own space (relativity)<br>c) Each point has its own time (relativity)<br>d) Finite space can only hold finite information |
| *Screen.* An expanding hyper-sphere surface:<br>a) Has three dimensions<br>b) Has no centre or edge<br>c) Spreads objects on it equally apart<br>d) An explosion on it ends up everywhere<br>e) Keeps vibrations on its surface | *Space.* Our expanding space:<br>a) Has three dimensions<br>b) Has no centre or edge<br>c) Is moving all the galaxies equally apart<br>d) Cosmic background radiation is everywhere<br>e) Limits entities to move within it |
| *Channels.* Grid node transfers:<br>a) Can simulate straight lines (geodesics)<br>b) Can change with load differentials | *Directions.* Objects naturally move in:<br>a) Straight lines (geodesics)<br>b) Geodesics are affected by gravity |
| *Time is processing*: Virtual time:<br>a) Is measured by processing events<br>b) Slows down with processing load<br>c) Is a discrete sequence of choices<br>d) Is one-way if a choice is irreversible<br>e) Can run in reverse for an isolated entity | *Time is relative.* Our time:<br>a) Is measured by atomic events<br>b) Dilates for high speed or by a large mass<br>c) Is simulated by movie state sequences<br>d) Quantum collapse is an irreversible choice<br>e) Anti-matter runs time in reverse |
| *The big bubble.* In the initial "rip":<br>a) One node split gives a sphere not a point<br>b) The chain reaction splits other nodes<br>c) Is a one-time creation of free information<br>d) Expands to dilute the waves on it | *The big bang.* The initial "bang":<br>a) Didn't give a massive black hole<br>b) The initial inflation was faster than light<br>c) The universe's information is constant<br>d) Cosmic background radiation is now cold |

22/24

# ISSUES

The following discussion questions highlight some of the issues covered. As no-one really knows, there are no absolute right or wrong answers. The aim is just to stimulate discussion:

1. If the physical world is a simulation, what is the image and what is the screen?
2. If the physical world is an image, what is its resolution and refresh rate?
3. Why can't the ongoing flux of our world ever stop?
4. Could an extra dimension wrap around our space?
5. Is space something or nothing?
6. Would a network simulating our universe be centralized or distributed?
7. How is a hyper-sphere surface like our space?
8. If movement is a grid transfer, what are "straight lines" in network terms?
9. If our time is virtual, how can we know if it slows down?
10. If time is a sequence of choices, can we redo the choices backwards, i.e. roll-back time?
11. How can time "flow backwards" for anti-matter?
12. How can space itself expand?
13. If space is expanding, what is it expanding into?
14. Why didn't the big bang immediately form a black hole?
15. Why is cosmic background radiation from the initial event still all around us?
16. If the total information of the universe is constant, how did it first arise?
17. How can a distributed network with no memory buffers solve synchronization problems?
18. If empty space is full, what fills it?

# REFERENCES

[1] Whitworth, B., "The emergence of the physical world from information processing," *Quantum Biosystems*, vol. 2, no. 1, pp. 221-249, 2010.

[2] M. Tegmark, "The Mathematical Universe," in *Visions of Discovery: Shedding New Light on Physics and Cosmology*, R. Chiao, Ed. Cambridge: Cambridge Univ. Press, 2007.

[3] L. Smolin, *The Trouble with Physics*. New York: Houghton Mifflin Company, 2006.

[4] K. Zuse, *Calculating Space*. Cambridge Mass.: MIT, 1969.

[5] E. Fredkin, "A Computing Architecture for Physics," in *Computing Frontiers 2005*, pp. 273-279, 2005.

[6] F. Wilczek, *The Lightness of Being: Mass, Ether and the Unification of forces*. New York: Basic Books, 2008.

[7] B. D'Espagnat, *Veiled Reality: An analysis of present-day quantum mechanical concepts*. Reading, Mass: Addison-Wesley Pub. Co., 1995.

[8] T. W. Campbell, *My Big TOE*, vol. 3. Lightening Strike Books, 2003.

[9] J. Barbour, *The End of Time: The next revolution in physics*. Oxford: Oxford University Press, 1999.

[10] C. E. Shannon and W. Weaver, *The Mathematical Theory of Communication*. Urbana: University of Illinois Press, 1949.

[11] G. McCabe, "Universe creation on a computer," 2004. .

[12] J. Mazur, *Zeno's Paradox*. London: Penguin Books, 2008.

*Simulating space and time, Brian Whitworth*
[13] J. D. Barrow, *New theories of everything*. Oxford: Oxford University Press, 2007.

[14] L. Randall, *Warped Passages: Unraveling the mysteries of the universe's higher dimensions*. New York: Harper Perennial, 2005.

[15] T. Berners-Lee, *Weaving The Web: The original design and ultimate destiny of the world wide web*. New York: Harper-Collins, 2000.

[16] P. Davies and J. R. Brown, *The Ghost in the Atom*. Cambridge: Cambridge University Press, 1999.

[17] L. Smolin, "Atoms of Space and Time," *Scientific American Special Issue, A Matter of Time*, pp. 56-65, 2006.

[18] P. Davies, "Emergent Biological Principles and the Computational Properties of the Universe," *Complexity*, vol. 10, no. 2, pp. 11-15, 2004.

[19] B. Greene, *The Fabric of the Cosmos*. New York: Vintage Books, 2004.

[20] J. Ambjorn, J. Jurkiewicz, and R. Loll, "The Self-Organizing Quantum Universe," *Scientific American*, vol. 299, no. 1, pp. 24-31, 2008.

[21] L. Smolin, *Three Roads to Quantum Gravity*. New York: Basic Books, 2001.

[22] S. Wolfram, *A New Kind of Science*. Wolfram Media, 2002.

[23] J. Case, D. S. Rajan, and A. M. Shende, "Lattice computers for approximating euclidean space," *Journal of the ACM*, vol. 48, no. 1, pp. 110-144, 2001.

[24] R. Penrose, "On the nature of quantum geometry," in *Magic Without Magic*, J. Klauder, Ed. San Francisco: Freeman, 1972, pp. 334-354.

[25] Edwin Abbott, "Flatland: a romance of many dimensions," *Project Gutenberg*, 1884. [Online]. Available: http://www.gutenberg.org/etext/201. [Accessed: 22-Feb-2010].

[26] S. Hawking and S. Penrose, *The nature of space and time*. Princeton, NJ.: Princeton University Press, 1996.

[27] G. P. Collins, "Computing with quantum knots," *Scientific American*, pp. 56-63, 2006.

[28] S. W. Hawking and J. B. Hartle, "The basis for quantum cosmology and Euclidean quantum gravity," *Phys. Rev.*, vol. 28, no. 2960, 1983.

[29] R. Dawkins, *The Selfish Gene*, vol. 2. Oxford University Press, 1989.

[30] Heylighen, Francis and Chielens, K., "Evolution of Culture, Memetics," in *Encyclopedia of Complexity and Systems Science*, Meyers, B, Ed. Springer, 2009.

[31] B. Whitworth and A. deMoor, "Legitimate by design: Towards trusted virtual community environments," *Behaviour & Information Technology*, vol. 22, no. 1, pp. 31-51, 2003.

[32] S. Kauffman and L. Smolin, "A possible solution to the problem of time in quantum cosmology," *arXiv Preprint*, vol. 9703026, pp. 1-15, 1997.

[33] J. B. Hartle, "The Physics of 'Now'," *Am.J.Phys.*, vol. 73, pp. 101-109, avail at http://arxiv.org/abs/gr-qc/0403001, 2005.

[34] E. H. Walker, *The Physics of Consciousness*. New York: Perseus Publishing, 2000.

[35] A. Guth, *The Inflationary Universe: The Quest for a New Theory of Cosmic Origins*. Perseus Books, 1998.

[36] D. Bohm, *Wholeness and the Implicate Order*. New York: Routledge and Kegan Paul, 1980.

[37] K. C. Cole, *The hole in the universe*. New York: Harcourt Inc, 2001.

[38] A. Einstein, "Ether and the Theory of Relativity," in *Address delivered on May 5th, 1920, in the University of Leyden, see http://www.tu-harburg.de/rzt/rzt/it/Ether.html*, .

[39] A. Watson, *The Quantum Quark*. Cambridge: Cambridge University Press, 2004.